%% file: LLM4OWL.tex
\documentclass[conference]{IEEEtran}
\usepackage{cite}
\usepackage{amsmath,amssymb,amsfonts}
\usepackage{algorithmic}
\usepackage[bookmarks=false]{hyperref}
\usepackage{listings}
\usepackage{graphicx}
\usepackage{textcomp}
\usepackage{xcolor}
\usepackage{tabularx}
\usepackage{booktabs}
\usepackage{tikz}
\usepackage{makecell}
\usepackage{svg}
\usepackage{caption}
\usepackage{subcaption}
\usepackage{lipsum}
\usepackage{comment}
\usepackage[numbers]{natbib}

\usepackage[nolist]{acronym}
\usetikzlibrary{shapes,snakes}
\usetikzlibrary{calc} 
\usetikzlibrary{arrows.meta}
\usetikzlibrary{positioning}
\usetikzlibrary{shapes, arrows}
\usetikzlibrary{arrows.meta,shapes.geometric,calc}
\usepackage{float}

\tikzset{
	>=Latex,
	line/.style={draw,->>},
	anode/.style={rectangle,draw,
		align=center,rounded corners,minimum height=4em,font=\strut},
	bnode/.style={anode,fill=white, font=\strut},
	cnode/.style={anode, fill=cyan!20, font=\strut},
treenode/.style = {circle,
	draw=black,thick, fill=white, align=center, minimum size=1cm},
root/.style     = {treenode, font=\footnotesize},
env/.style      = {treenode, font=\footnotesize}, 
dummy/.style    = {circle,draw}
	
}
\tikzstyle{decision} = [diamond, draw, fill=yellow!20, 
    text width=6em, text badly centered, node distance=3cm, inner sep=0pt, line width=0.8pt, minimum height=9em,
    minimum width=9em]
\tikzstyle{block} = [rectangle, draw, fill=cyan!20, 
    text width=6.7em, text centered, rounded corners, minimum height=4em, line width=0.8pt]
\tikzstyle{line} = [draw, -latex, line width=0.8pt]
\tikzstyle{cloud} = [draw, ellipse,fill=red!25, node distance=3cm, minimum height=2em]
\tikzstyle{method} =  [trapezium, trapezium left angle=70, trapezium right angle=-70,text centered,text width = 2cm,minimum height=1cm, minimum width=2cm, draw=black, fill=green!20, line width=0.8pt]

\def\BibTeX{{\rm B\kern-.05em{\sc i\kern-.025em b}\kern-.08em
    T\kern-.1667em\lower.7ex\hbox{E}\kern-.125emX}}

\definecolor{light-gray}{gray}{0.95}
\lstdefinestyle{mystyle}{
    backgroundcolor=\color{light-gray},  
    basicstyle=\ttfamily \footnotesize,
    breaklines=true
}

\begin{document}
\bstctlcite{IEEEexample:BSTcontrol}
\begin{acronym}

  \acro{llm}[LLM]{\textit{Large Language Model}}
  \acroplural{llm}[LLMs]{\textit{Large Language Models}}

  \acro{odp}[ODP]{\textit{Ontology Design Pattern}}
  \acroplural{odp}[ODPs]{\textit{Ontology Design Patterns}}
  
  \acro{nlp}[NLP]{\textit{Natural Language Processing}}

  \acro{nl}[NL]{\textit{Natural Language}}

  \acro{rnn}[RNN]{\textit{Recurrent Neural Network}}
  \acroplural{rnn}[RNNs]{\textit{Recurrent Neural Networks}}
  
  \acro{sparql}[SPARQL] {\textit{SPARQL Protocol and RDF Query Language}}

  \acro{http}[HTTP]{\textit{Hypertext Transfer Protocol}}

  \acro{kg}[KG]{\textit{Knowledge Graph}} 
  \acroplural{kg}[KGs]{\textit{Knowledge Graphs}}
  
  \acro{owl}[OWL]{\textit{Web Ontology Language}}

  \acro{cps}[CPS]{\textit{Cyber-Physical System}}
  \acroplural{cps}[CPS]{\textit{Cyber-Physical Systems}}

  \acro{cq}[CQ]{\textit{Competency Question}}
  \acroplural{cq}[CQs]{\textit{Competency Questions}}

  \acro{uri}[URI]{\textit{Uniform Resource Identifier}}
  \acroplural{uri}[URIs]{\textit{Uniform Resource Identifiers}}

  \acro{scq}[SCQ]{\textit{standard-compliant question}}
  \acroplural{scq}[SCQs]{\textit{standard-compliant questions}}

  \acro{nscq}[NSCQ]{\textit{non-standard-compliant question}}
  \acroplural{nscq}[NSCQs]{\textit{non-standard-compliant questions}}
\end{acronym}

\title{Chatbot-Based Ontology Interaction Using Large Language Models and Domain-Specific Standards}

\author{
    \IEEEauthorblockN{
        Jonathan Reif\IEEEauthorrefmark{1},
        Tom Jeleniewski\IEEEauthorrefmark{1},
        Milapji Singh Gill\IEEEauthorrefmark{1},
        Felix Gehlhoff\IEEEauthorrefmark{1},
        Alexander Fay\IEEEauthorrefmark{2}%
    } 
    
    \IEEEauthorblockA{
        \IEEEauthorrefmark{1}Institute of Automation Technology\\
        \textit{Helmut Schmidt University Hamburg, Germany}\\
        {\small \{jonathan.reif, tom.jeleniewski, milapji.gill, felix.gehlhoff\}@hsu-hh.de}\\
        \IEEEauthorrefmark{2}Chair of Automation\\
        \textit{Ruhr University, Bochum, Germany}\\
        {\small alexander.fay@rub.de}\\
       }
       \vspace{-1cm}
}

\maketitle

\begin{abstract}
The following contribution introduces a concept that employs Large Language Models (LLMs) and a chatbot interface to enhance SPARQL query generation for ontologies, thereby facilitating intuitive access to formalized knowledge. 
Utilizing natural language inputs, the system converts user inquiries into accurate SPARQL queries that strictly query the factual content of the ontology, effectively preventing misinformation or fabrication by the LLM. 
To enhance the quality and precision of outcomes, additional textual information from established domain-specific standards is integrated into the ontology for precise descriptions of its concepts and relationships. 
An experimental study assesses the accuracy of generated SPARQL queries, revealing significant benefits of using LLMs for querying ontologies and highlighting areas for future research. 
\end{abstract}


\input{01_Introduction}

\input{02_Background}

\input{03_Concept}

\input{04_Summary_and_Outlook}

\section*{Acknowledgment}
This research [project ProMoDi and LaiLa] is funded by dtec.bw – Digitalization and Technology Research Center of the Bundeswehr. dtec.bw is funded by the European Union – NextGenerationEU.

\bibliographystyle{IEEEtranN}

\end{document}

%% file: 01_Introduction.tex
\section{Introduction}
\label{sec:introduction}
In the context of \acp{cps} and Industry 4.0, domain-specific ontologies serve as highly advantageous knowledge bases. 
Ontologies enable efficient data integration and enhance the interoperability of different \acp{cps} by establishing clear semantics and formalizing domain knowledge \cite{Hildebrandt.2020}. 
These attributes are particularly beneficial in industrial applications where data is dispersed across various heterogeneous sources. 
Thus, ontologies are essential for managing the increasing complexity of \ac{cps} engineering, operation, and maintenance. 
Particularly, their use within assistance systems in various Industry 4.0 application scenarios improves decision-making and optimizes processes across different life cycle phases of \ac{cps} \cite{Jeleniewski.SemanticModel2023, Reif.2023, Gill.2023}.

However, significant challenges exist in facilitating interaction between end users and ontologies. 
Ontologies are inherently complex and not easily understandable, making them difficult for end users to engage with effectively \cite{Chen.2022}.
Traditionally, knowledge querying is conducted based on predefined \acp{cq} and static \ac{sparql} queries \cite{Hildebrandt.2020}. 
These are neither intuitive for someone without Semantic Web expertise nor user-friendly. 
Especially in view of the lack of ontology experts, this problem is even more critical \cite{Hildebrandt.2020}. 
Many users find it difficult to formulate complex \ac{sparql} queries. This significantly limits the flexibility and efficiency of knowledge retrieval.
In this context, the communication between a domain-specific ontology and an end user can be substantially improved by employing chatbots and \acp{llm} \cite{Chen.2022, Avila.2024}.

Currently, \ac{llm}-based assistance system experience increased significance, driven by the latest accomplishments in the \ac{llm} community. 
However, relying solely on \ac{llm}-based approaches poses significant risks.
Due to the lack of traceability of \acp{llm} and their potentially highly creative interpretative abilities, it cannot be ensured that the \ac{llm} extracts an answer directly from a credible source instead of creating its own probability-based responses.
As a result, incorrect information could be conveyed to the user, potentially causing errors. This poses a significant threat, especially in industrial applications, where the accuracy and correctness of answers are crucial, and hallucinated responses generated by an \ac{llm} could have severe consequences. 
Therefore, it is beneficial to adopt an approach that combines the advantages of ontologies — the formal, structured provision of factual knowledge — with those of \acp{llm}, which offer intuitive and easy use. For this reason, we propose a concept to improve the process of automated \ac{sparql} query generation by leveraging \acp{llm} and information from domain-specific standards. This approach aims to enhance user-friendliness by providing an intuitive interface for querying complex ontologies. 

The paper is structured as follows: In Sec.~\ref{sec:RelatedWork}, we introduce the requirements for the proposed concept and analyze related work concerning these requirements. 
Sec.~\ref{sec:Concept} details the concept. Preliminary results from an initial experimental study are presented in Sec.~\ref{sec:results}. 
Finally, Sec.~\ref{sec:conclusion} summarizes the contribution and depicts future research directions.

%% file: 02_Background.tex
\section{Requirements and Related Work}  \label{sec:RelatedWork}

\subsection{Requirements}
\noindent \textbf{R1: Intuitive communication between the user and the individually created ontology}\\
As described in Sec.~\ref{sec:introduction} the use of ontologies can pose significant challenges for non-experts. 
Therefore, one of the key requirements is to lower the usage barrier by assisting the user in their communication with the ontology. 
This involves creating an interface or a system that is user-friendly and intuitive, allowing users to interact with the ontology in a way that feels natural and straightforward. 
Thus, users should be permitted to formulate queries using their own words, rather than being required to use specific terms or codes \cite{Yani.2021}.

\noindent \textbf{R2: Flexible queries to the ontology based on the user's current information requirements}\\
Given the need to facilitate the user’s interaction with the ontology, the second requirement is the ability to generate customized queries. 
These queries should be flexible and adapt to the user’s current information needs.  
This flexibility enhances the user-friendliness of the system and ensures that the user can extract the most relevant and useful information from the ontology \cite{SaiSharath.2021}.

\noindent \textbf{R3: Accuracy and traceability of answers}\\
While making the ontology user-friendly is important, it would be futile if the information retrieved is not accurate \cite{MartinezGil.2022}.
Therefore, the third requirement is to guarantee the accuracy of the answers provided by the assistance-system.
This requirement becomes even more critical in an industrial setting where the accuracy of information is paramount.
In this context, the correctness of the answers additionally has to be traceable by users to ensure correctness as inaccurate information can lead to costly mistakes, inefficiencies, and even safety risks. 
Therefore, the system must be designed to provide precise, reliable, and traceable answers. 

\subsection{Related Work}
\citet{Chen.2021} present a semantic embedding framework for \ac{owl} ontologies.
It utilizes a combination of random walks and word embedding techniques to encode the semantics of \ac{owl} ontologies by considering their graph structure, lexical information, and logical constructors.
The results suggest good accuracy. 
However, because the generation of answers is solely based on the use of an \ac{llm}, there is no traceability of the generated answers.

\citet{Chen.2022} introduce a system designed to efficiently generate \ac{sparql} queries for so-called question answering systems. 
The primary objective of the system is to reduce query costs while maintaining high accuracy in generating \ac{sparql} queries, which are used to retrieve answers from databases. 
The approach employs a \ac{rnn} to generate \ac{sparql} queries from learned and labeled keywords. 
Building on this, \citet{Chen.2023} describe the enhancement of question answering systems through advanced \ac{nlp} techniques and multi-label classification also using \acp{rnn}.
They emphasize the use of \ac{nlp} to interpret and process user queries in \ac{nl}, converting them into a format that can be used to generate \ac{sparql} queries effectively. 
This involves the use of technologies such as tokenization, lemmatization, and part-of-speech tagging to understand the semantic structure of the queries.
Although both publications also pursue the idea of translating user questions into \ac{sparql} queries, they refrain from using \acp{llm} \cite{Chen.2022, Chen.2023}.
However, \citet{Chen.2022} also emphasize that when converting a user question into query grammar, the accuracy of the generation of the query decides whether an answering system can provide a correct answer.

\citet{Avila.2024} conducted preliminary experiments to evaluate \textit{ChatGPT's} (GPT-3.5) ability to answer \ac{nl} questions over a \ac{kg} in domains such as families, people, and jobs. 
Various setups were tested, including direct answering and text-to-SPARQL translation using either or both the TBox and ABox of the \ac{kg}. 
The results indicated that the text-to-SPARQL approach utilizing both the TBox and ABox yielded the best performance.
Furthermore, \citet{Avila.2024} present a framework designed to optimize the translation of \ac{nl} questions into \ac{sparql} queries. 
It operates in two stages: an offline stage that generates indices mapping \ac{kg} terms (TBox and ABox) to their \acp{uri}, and an online stage that uses these indices to translate \ac{nl} questions into SPARQL queries and generate responses. 
By reducing the number of tokens processed, the framework decreases the likelihood of hallucinations and enhances support for large \acp{kg}.
However, the effect of providing explanations of the graph to the \ac{llm} as well as the level of complexity to which the \ac{llm} can generate \ac{sparql} queries in an reliable manner where not examined.

%% file: 03_Concept.tex
\section{Concept for Chatbot-Based Interaction with Ontologies} \label{sec:Concept}
In the following, a concept is introduced that utilizes a chatbot-based interface for user interactions with ontologies, providing flexible querying options.
As mentioned in Sec.~\ref{sec:RelatedWork}, previous approaches utilize \acp{llm} directly as a querying medium. 
However, this is associated with significant risks in industrial settings as explained in Sec.~\ref{sec:introduction}.
Therefore, our approach is based on maintaining \ac{sparql}-based querying.

\begin{figure}[H]
  \centering
  \includegraphics[width=0.88\columnwidth]{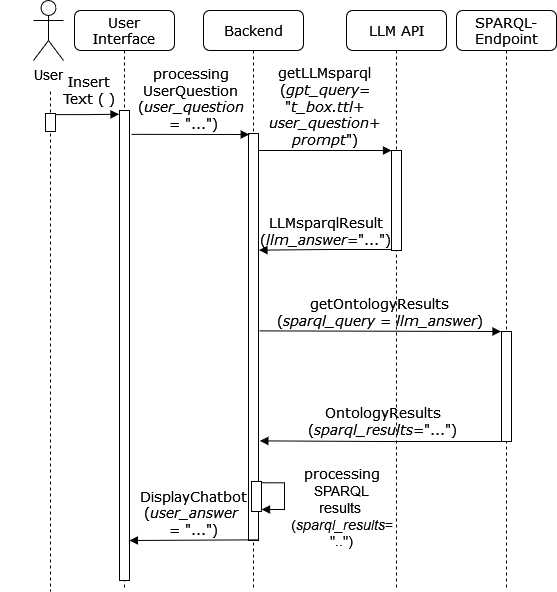}
  \caption{Concept for \ac{llm}-based interaction with ontologies}
  \label{fig:concept}
\end{figure}

To support domain-specific experts with limited expertise in handling Semantic Web technologies, \acp{llm} are used in this approach to generate SPARQL queries. 
The approach is illustrated in Fig.~\ref{fig:concept}.
Users interact with the ontology via a chat interface, allowing them to pose questions that are processed in the backend.
These queries are sent to the \ac{llm} through the \textit{ChatGPT API}, incorporating predefined prompts, as explained in detail in Sec.~\ref{subsubsec:prompts} and the TBox.
By including only the TBox in the prompt, the security advantage is that the explicit ABox knowledge stays separate from the \ac{llm} and can only be accessed through \ac{sparql} queries. 
This approach ensures that sensitive industrial information remains protected within the company’s internal systems.
However, the \ac{llm} does not directly answer the input question. 
Instead, it uses the predefined prompt to transform the question into a \ac{sparql} query.
This query is then returned to the backend and executed at the \ac{sparql} endpoint. 
The results are processed in the backend and displayed to the user through the user interface, providing the answers required.

This approach ensures that the \ac{llm} produces no incorrect or fabricated answers, as only "true" results that are embedded within the model are retrieved without invention or interpretation. 
However, while the responses are verifiable because they reflect the model's actual knowledge, they are not fully validated. 
Errors may still persist in the generated queries, leading to answers that do not match the original question or, in some cases, no answers being returned at all.

Considering that potential users may not be familiar with the terminology used in the ontology, it is crucial to accommodate their ability to accurately formulate questions via the chat interface.
Special attention should be given to the following aspects to ensure effective user interaction:

\subsubsection{Prompts} \label{subsubsec:prompts}
Given that neither the user nor the \ac{llm} is aware of the TBox associated with the ontology to be queried, it is essential to supplement the chat-based questions with prompts that incorporate the required contextual knowledge.
These prompts must comprehensively detail the TBox, encapsulating the modeling concepts and terminologies of the ontology.
By integrating extensive descriptions of the ontology, including its classes, properties, and relations, within the prompts, the \ac{llm} can accurately interpret user questions and transform them into precise \ac{sparql} queries. 
For this reason, the TBox containing the modeling concept should be transferred together with the user question.
These prompts including the TBox serve as a guide for translating domain-specific knowledge into executable \ac{sparql} queries, thus ensuring that the \ac{llm} grasps the necessary context and specifics for accurate query formulation.

\subsubsection{Creating Ontology Design Patterns}
In the industrial domain, fixed terminologies defined by industry standards are prevalent. 
For building ontologies, we employ \acp{odp} that adhere to these established standards, functioning as templates for creating knowledge graphs.
\citet{Hildebrandt.2020} outline a methodological approach for building ontologies based on \acp{odp}.
When these aligned \acp{odp}, resembling a TBox for the problem domain context, accompany the chat-based queries, the \ac{llm} can examine the structures and terminology to accurately translate the chat query into \ac{sparql} in accordance with the \acp{odp}.

Enhancing \acp{odp} with \texttt{rdfs:comment} annotations is critical, as these provide additional context about the classes, object properties, and data properties. 
This extra layer of context assists the \ac{llm} in disambiguating terms that may be ambiguous or have multiple interpretations based on their literal meanings. 
By leveraging \texttt{rdfs:comment}, the \ac{llm} gains deeper insights into the intended semantics of the ontology elements, thereby enhancing its capability to accurately transform user questions into \ac{sparql} queries. 
This strategy ensures that the generated queries are more closely aligned with the underlying ontology, minimizing misinterpretations and enhancing the reliability of the \ac{sparql} queries.

\section{Preliminary Results}
\label{sec:results}
Experiments were conducted to assess the ability of \acp{llm} to generate \ac{sparql} queries consistent with the criteria outlined in Sec.~\ref{sec:RelatedWork}.
The study investigated factors affecting the quality of these queries to evaluate the practical use of \acp{llm} in this domain.
\textit{ChatGPT-4o} was employed to create \ac{sparql} queries for various \acp{odp}, utilizing prompts that included specific \ac{odp} information in plain text and a corresponding question that the query aimed to answer.

The \acp{odp} used in the study were VDI 3682 (\textit{Formalized Process Description}) \cite{VDI3682}, DIN EN 61360 (\textit{Property Descriptions of Technical Systems}) \cite{DINEN613601}, and VDI 2206 (\textit{Descriptions of Machine Structures}) \cite{VDIVDE2206}.
Questions were crafted in two styles to assess the influence of phrasing on query quality: Firstly, \acp{scq} were posed, ensuring that the terminology adhered to established standards. Secondly, \acp{nscq} were formulated, incorporating more generic terminology typical for a non-expert.

Additionally, these questions were posed using \acp{odp} both with and without annotations in \texttt{rdfs:comment}, aiming to determine if such comments enhance query quality.
\begin{table}[H]
\caption{Examined question categories according to \cite{Rony.2022}}
    \begin{minipage}{0.1\linewidth}
        \centering
        \begin{tikzpicture}[scale=0.32]
            \draw[fill=gray!10] (1,14) -- (2,-5) -- (0,-5) -- cycle;
            \node[rotate=90] at (1,0) {{Complexity}};
        \end{tikzpicture}
    \end{minipage}
    \hfill
    \begin{minipage}{0.9\linewidth}
        \centering
        \begin{tabularx}{\linewidth}{lXr}
            \toprule 
            \footnotesize
            \textbf{Category} & \textbf{SCQ Example} & \textbf{ODP}
            \\
            \toprule 
            Boolean & \textit{Is the sensor part of a module in the system?} & VDI 2206\\
            \midrule  
            Count & \textit{How many technical resources are contained within the system?} & VDI 3682\\
            \midrule
            Rank & \textit{Could you provide the values contained in the model in ascending order?} & DIN EN 61360
            \\
            \midrule
            Simple & \textit{Which process operators are performed in process X?}& VDI 3682\\
            \midrule
            String & \textit{Is there a DataElement with the name "ResultAccuracy"?}& DIN EN 61360\\
            \midrule
            Two Hop & \textit{Which components are part of a module and which system includes this module?}& VDI 2206\\
            \midrule
            Two Intent & \textit{Of which process operators is process X composed? To which technical resources are these process operators assigned?}& VDI 3682\\
            \bottomrule
        \end{tabularx}
        
        \label{tab:questions}
    \end{minipage}
\end{table}
The complexity of questions was graded into seven categories according to the scheme proposed by \citet{Rony.2022}, shown in Tab.~\ref{tab:questions}. \textit{Boolean}, \textit{Count}, and \textit{Rank} represent simpler queries to the ontology, intended to yield a True/False outcome, a numerical count, or a ranking, respectively. \textit{Simple}, \textit{String}, and \textit{Two Hop} require querying more complex graph relationships or more specific words, necessitating a greater semantic understanding. \textit{Two Intent} is the most complex category, as it essentially requires two responses and the merging of multiple triples. For each category, an example with a SCQ is listed in Tab.~\ref{tab:questions}, along with the corresponding ODP. Overall, the experimental study encompasses 28 questions for each ODP, resulting in a total of 84 questions with analyzed results.

Preliminary results indicate a generally good performance of the tested \ac{llm} in generating \ac{sparql} queries.
In Tab.~\ref{tab:results}, the questions categories, exhibiting similar patterns in the results, were classified into three clusters.
The results show that simpler questions (\textit{Boolean}, \textit{Count}, \textit{Rank}) generally yielded more accurate \ac{sparql} queries, regardless of the presence of \texttt{rdfs:comment} in the tested ontologies.  
Moreover it was found that precise \ac{scq} formulation significantly affected query quality.
However, for more complex questions (e.g., \textit{Two Intent}), \textit{ChatGPT-4o} often generated inaccurate or imprecise queries, typically failing to identify the correct instance. 
The results also suggest that \acp{odp} augmented with \texttt{rdfs:comment} produced queries with greater precision, supporting the hypothesis that detailed comments in ontologies positively impact automated \ac{sparql} query generation.
Accordingly, it can be concluded that annotations enhance not only human comprehension of ontologies but also offer significant benefits for \acp{llm} in terms of query generation accuracy and effectiveness.
Overall the suitability of \acp{llm} for (automated) \ac{sparql} query generation could be shown. 

\begin{table}[H]
\caption{Preliminary Results: Percentage of correctly generated SPARQL queries}
        \centering
        \begin{tabular}{l|cccc}
            \toprule 
            & \multicolumn{2}{c}{\textbf{w/o comments}} & \multicolumn{2}{c}{\textbf{commented}} \\
            \cmidrule(r){2-3} \cmidrule(r){4-5}
            \textbf{Categories} & SCQs & NSCQs & SCQs & NSCQs \\
            \toprule
            Boolean, Count, Rank & 100\% & 100\% & 100\% & 100\% \\
            \midrule
             Simple, String, Two Hop & 89\% & 44\% & 100\% & 78\% \\
            \midrule
             Two Intent & 67\% & 0\% & 100\% & 67\%\\
            \bottomrule
        \end{tabular}
\label{tab:results}
\end{table}

%% file: 04_Summary_and_Outlook.tex
\section{Summary and Future Work} \label{sec:conclusion}
This paper investigates the suitability of \acp{llm} for generating \ac{sparql} queries to simplify user interaction with ontologies. 
By combining the intuitive usability of \ac{llm}-based chat applications with the formal, structured knowledge provision of ontologies, the proposed concept is particularly beneficial in an industrial context.
An experimental study using \textit{ChatGPT-4o} assessed the accuracy of \ac{sparql} queries generated under various conditions and highlighted the value of incorporating \texttt{rdfs:comment}. 

Future evaluation will focus on assessing the overall concept and exploring which \acp{llm} are best suited for the task.
Developing strategies to reduce errors in \ac{sparql} query generation, especially for complex queries, is crucial for enhancing accuracy and reliability.
Improving user interaction with ontology-based systems by refining prompts and optimizing how the ontology context is provided to the language model is also important. 
Additionally, exploring the impact of detailed \texttt{rdfs:comment} on the quality of generated \ac{sparql} queries is needed, including testing with more complex ontologies and varying the detail level of \texttt{rdfs:comment}. 
Implementing robust validation mechanisms is essential to ensure the accuracy and trustworthiness of the generated queries, especially in industrial settings where incorrect information can have significant consequences.